\documentclass[12pt]{article}
\usepackage[dvips]{graphics}

\catcode`@=11
\def\@citex[#1]#2{\if@filesw\immediate\write\@auxout{\string\citation{#2}}\fi
  \def\@citea{}\@cite{\@for\@citeb:=#2\do
    {\@citea\def\@citea{,\penalty\@m}\@ifundefined
      {b@\@citeb}{{\bf ?}\@warning
       {Citation `\@citeb' on page \thepage \space undefined}}%
\hbox{\csname b@\@citeb\endcsname}}}{#1}}

\def\citer{\@ifnextchar
[{\@tempswatrue\@citexr}{\@tempswafalse\@citexr[]}}
\def\@citexr[#1]#2{\if@filesw\immediate\write\@auxout{\string\citation{#2}}\fi
  \def\@citea{}\@cite{\@for\@citeb:=#2\do
    {\@citea\def\@citea{--\penalty\@m}\@ifundefined
       {b@\@citeb}{{\bf ?}\@warning
       {Citation `\@citeb' on page \thepage \space undefined}}%
\hbox{\csname b@\@citeb\endcsname}}}{#1}}
\catcode`@=12

\newcommand{\lsim}{\buildrel < \over {_\sim}}

\newcommand{\be}{\begin{equation}}
\newcommand{\ee}{\end{equation}}
\newcommand{\bea}{\begin{eqnarray}}
\newcommand{\eea}{\end{eqnarray}}
\newcommand{\ba}{\begin{array}}
\newcommand{\ea}{\end{array}}
\newcommand{\ovl}{\overline}
\newcommand{\uonep}{$U(1)^\prime${ }}
\newcommand{\Tr}{{\rm Tr}}
\newcommand{\integer}{{\sf Z \hspace{-5pt} Z}}

\newcommand{\plb}[2]{{\em Phys. Lett.}              {\bf #1B}, #2 }
\newcommand{\npb}[2]{{\em Nucl. Phys.}              {\bf B#1}, #2 }

\newcommand{\pr }[2]{{\em Phys. Rep.}               {\bf  #1}, #2 }

\newcommand{\prd}[2]{{\em Phys. Rev.}               {\bf D#1}, #2 }
\newcommand{\prl}[2]{{\em Phys. Rev. Lett.}         {\bf  #1}, #2 }
\newcommand{\zpc}[2]{{\em Z. Phys.}                 {\bf C#1}, #2 }
\newcommand{\sci}[2]{{\em Science}                  {\bf  #1}, #2 }

\newcommand{\ijm}[2]{{\em Int. J. Mod. Phys.}       {\bf A#1}, #2 }

\newcommand{\epj}[2]{{\em Eur. Phys. J.}            {\bf C#1}, #2 }
\newcommand{\con}[2]{                               {\bf  #1}, #2 }
\newcommand{\etal}{{\em et al.}}
\newcommand{\ibid}{{\em ibid.}}

\renewcommand{\LARGE}{\large}

\begin{document}

\hfill \\ \vspace{10pt} 
\hfill {\normalsize UPR--886--T} \\
\vspace{20pt}
\centerline{\LARGE \bf Chiral Models of Weak Scale Supersymmetry}
\vspace{15pt}
\centerline{\large Jens Erler}
\centerline{
\it Department of Physics and Astronomy, University of Pennsylvania,}
\centerline{\it Philadelphia, PA 19104-6396}
\centerline{\it E-mail: erler@langacker.hep.upenn.edu}

\begin{abstract}
I discuss supersymmetric extensions of the Standard Model containing an extra
\uonep gauge symmetry which provide a solution to the $\mu$-problem and at 
the same time protect the proton from decaying via dimension~4 operators. 
Moreover, all fields are protected by chirality and supersymmetry from 
acquiring high scale masses. The additional requirements of anomaly 
cancellation and gauge coupling unification imply the existence of extra matter
multiplets and that several fields participate in \uonep symmetry breaking 
simultaneously. While there are several studies addressing subsets of these 
requirements, this work uncovers simultaneous solutions to all of them. It is 
surprising and encouraging that extending the Minimal Supersymmetric Standard 
Model by a simple $U(1)$ factor solves its major drawbacks with respect to 
the non-supersymmetric Standard Model, especially as current precision data 
seem to offer a hint to the existence of its corresponding $Z^\prime$ boson. 
It is also remarkable that there are many solutions where the \uonep charges of
the known quarks and leptons are predicted to be identical to those $E_6$ 
motivated $Z^\prime$ bosons which give the best fit to the data. I discuss
the solutions to these constraints including some phenomenological 
implications. 
\end{abstract}

\section{Introduction}
Besides its conceptual elegance, the prime motivation to consider low energy 
supersymmetry (SUSY)~\cite{Nilles84} is the stabilization of the electroweak 
scale under radiative corrections. However, the Minimal Supersymmetric Standard
Model (MSSM) provides no explanation to the $\mu$-problem~\cite{Kim84}, i.e., 
to why the scales of SUSY breaking and of the supersymmetric bilinear Higgs 
$\mu$-term,
\be
  \mu h_d h_u,
\ee
are of comparable magnitude. The MSSM should therefore be extended by at least 
some sector solving the $\mu$-problem. In this sense, the term MSSM is 
a misnomer. 

Moreover, in sharp contrast to the non-supersymmetric Standard Model (SM)
where the stability of the proton can be understood entirely by gauge 
invariance, the MSSM predicts rapid proton decay via dimension~4 operators, 
unless an additional discrete symmetry, such as $R$-parity, is imposed ad hoc. 
While this is a logical possibility, it severely compromises the initial 
elegance and one prefers to save the successful features of the SM when 
exploring its alternatives. It has also been argued~\cite{Font89} that those
discrete symmetries\footnote{Continuous global symmetries are not expected 
to arise from string theory~\cite{Banks88}.} which survive as remnants of some 
underlying string model, might either be unsuited to forbid (at least) 
the dangerous dimension~4 operators, or arise only along some fine-tuned flat 
scalar field direction. Indeed, baryon number conservation by the principles of
gauge invariance (and renormalizability) has been on the wish list of SUSY 
practitioners for almost two decades~\cite{Weinberg82}.

It is well known that either of the problems addressed in the previous two 
paragraphs can be solved by the introduction of an extra \uonep symmetry.
Examples for either case are reviewed in the next section using $U(1)$ 
groups motivated by $E_6$ Grand Unified Theories (GUTs). Additional $U(1)$s 
have long been considered as very well motivated extensions of 
the MSSM~\cite{Cvetic97}; they are predicted in most GUTs and appear copiously
in superstring theories. In this article, I address the question whether it is 
possible to find models in which an extra \uonep solves both problems 
{\em simultaneously}. 

If such an extension is free of gauge and gravitational anomalies, one can 
allow a grand desert scenario between the electroweak scale and a more 
fundamental high mass scale. This is desirable since our confidence in 
the MSSM is boosted by the observation of approximate gauge coupling 
unification~\cite{Dimopoulos81} (as predicted by the simplest and most economic
GUT and string models) at a scale somewhat below the Planck scale. An extra 
\uonep does not affect the renormalization group equations at the one-loop 
level, and it is conceivable that the quality of unification even improves at 
two-loop precision. In order to rigorously protect the electroweak scale from 
higher mass scales, I also demand that the field content of the models be 
completely chiral before gauge and supersymmetry breaking. I.e., all 
superfields should transform non-trivially with respect to the low-energy 
$SU(3)\times SU(2)\times U(1)_Y\times U(1)^\prime$ gauge symmetry, and all 
bilinear (mass) terms must be forbidden by gauge invariance. Note, that these 
requirements present a strong set of constraints, and it is not clear 
{\em a priori\/}, whether any solutions exist. The key ingredient necessary 
to arrive at a solution is that at least two MSSM singlet fields which are 
charged under the \uonep have to acquire vacuum expectation values (VEVs).

Electroweak precision data imply further hints and constraints. The predictions
of the SM are generally in very good agreement with experiments provided 
the Higgs boson mass, $M_H$, is smaller than about 
250~GeV~\cite{Erler99,Erler99A}. This is also predicted by the MSSM and many 
extensions ($M_H \lsim 150$~GeV)~\cite{Carena96}. Moreover, the superpartners 
and extra Higgs fields decouple from the precision observables over large parts
of the MSSM parameter space, so that the MSSM is likewise in agreement with 
observations~\cite{Erler98}. Addition of an extra $Z^\prime$ boson can even 
improve the global fit to all data~\cite{Erler00}, driven mostly by the parity
violation experiments in Cs~\cite{Wood97,Amaldi87}, but the results in 
Tl~\cite{Edwards95}, and from the $Z$ lineshape measurements at 
LEP~1~\cite{Mnich99} also play a role. Note, that these fits suggest 
a $Z^\prime$ boson mass $M_{Z^\prime} \lsim {\cal O} (\mbox{1 TeV})$ and 
continue to yield results on $M_H$ consistent with SUSY~\cite{Erler00}.

Fits to the oblique parameters, $S$, $T$, and $U$~\cite{Peskin90} (or 
$\epsilon_1$, $\epsilon_2$, and $\epsilon_3$~\cite{Altarelli90}), describing 
new physics contributions to vector boson self-energies, yield results 
consistent with zero~\cite{Erler99A} (for Higgs masses in the MSSM range). This
implies restrictions on extra matter fields which can contribute significantly 
to $S$ and $T$, but if they are non-chiral (with respect to the SM gauge group)
and approximately mass degenerate their effects are rendered small. If they
arise in complete representations of $SU(5)$, they do not affect gauge coupling
unification at the one-loop level, nor its scale. However, in the bottom-up
approach of this paper I do not assume an $SU(5)$-GUT symmetry. I will show
that other matter configurations can preserve unification, as well.

To reiterate, I will assume that there is an additional \uonep symmetry with 
a mechanism built in to solve the SUSY $\mu$-problem. I will classify 
the solutions to the conditions of anomaly cancellation, gauge coupling 
unification, and chirality. To avoid potential problems with fractionally 
charged states, I will also assume $SU(5)$-type charge quantization. About half
of the resulting models {\em predict\/} a sufficiently long proton lifetime. 

In the next section, I review the situation for the most popular \uonep 
extensions arising from $SO(10)$ and $E_6$ GUTs~\cite{Hewett89}. 
In Section~\ref{1singlet}, I introduce the anomaly conditions and show that it 
is not possible to find chiral models if only one MSSM singlet field 
develops a VEV. Section~\ref{2singlets} shows that there are 33 solutions 
(family universal with respect to the ordinary fermions) if there are two 
singlets acquiring VEVs in the course of \uonep breaking. Five of these models 
involve a number of ${\bf 5} + {\bf \ovl{5}}$ representations and singlets of 
$SU(5)$, but no extra matter beyond that. Allowing {\em incomplete} 
representations of $SU(5)$ and the weaker requirements of gauge unification and
absence of fractionally charged states yield the remaining 28 solutions. I will
also briefly address the prospects to derive these models from 
the $E_8 \times E_8$ heterotic string theory. Some of the phenomenological 
implications of the type of models obtained in Section~\ref{2singlets} are 
discussed in Section~\ref{pheno}. My conclusions are presented in 
Section~\ref{con}.

\section{$E_6$ inspired \uonep models}
\label{E6example}
Consider, as an illustrative example, the case of an $E_6$ gauge group which is
free of anomalies in all its representations. A fundamental {\bf 27} 
representation contains a pair of $SU(2)$ doublets, $h_d$ and $h_u$, carrying 
the appropriate $U(1)_Y$ hypercharges to serve as the MSSM Higgs doublets, 
i.e., allowing the Yukawa couplings,
\be
   q h_d \ovl{d},       \hspace{20pt} 
   q h_u \ovl{u},       \hspace{20pt} 
   l h_d e^+,           \hspace{20pt}
   l h_u \ovl{\nu}.
\label{Yukawa}
\ee
Here $l$ and $q$ are the lepton and quark doublet superfields\footnote{I denote
the MSSM multiplets by lower case letters, while additional fields will be 
capitalized.}, $e^+$, $\ovl{d}$, and $\ovl{u}$ denote the $SU(2)$ singlets, and
$\ovl{\nu}$ refers to the right-handed neutrino superfield. If one identifies 
the \uonep with the $U(1)_\psi$ defined by $E_6 \rightarrow SO(10) \times 
U(1)_\psi$, then $h_d$ and $h_u$ have equal \uonep charges, excluding an 
elementary $\mu$-term. A {\bf 27} also includes an MSSM singlet superfield, 
$S$, which has the right \uonep charge to allow a trilinear Yukawa term,
\be
\label{SHH}
   S h_d h_u,
\ee
in the superpotential. If the scalar component of $S$ develops a VEV triggered 
by SUSY breaking, it breaks the \uonep gauge symmetry and induces an effective 
$\mu$-term, linking these scales. Since it is assumed that electroweak symmetry
breaking is triggered by SUSY breaking, as well, all scales are related and 
the $\mu$-problem is solved~\cite{Suematsu95,Cvetic96}. Moreover, one predicts 
$M_{Z^\prime} \lsim {\cal O} (\mbox{1 TeV})$ if one wants to avoid excessive 
fine-tuning. The $U(1)_\psi$ symmetry also forbids all baryon number, {\bf B}, 
and lepton number, {\bf L}, violating terms in the renormalizable
superpotential of the MSSM,
\be
\label{LLE}
   l h_u,       \hspace{20pt} 
   l l e^+,     \hspace{20pt} 
   l q \ovl{d}, \hspace{20pt} 
   \ovl{d} \ovl{d} \ovl{u}.
\ee
However, each {\bf 27} also contains an extra pair of $SU(2)$ singlet quarks, 
$D$ and $\ovl{D}$. Upon \uonep breaking they are expected to receive TeV scale 
masses through terms of the form\footnote{Terms of this type can also trigger
radiative \uonep symmetry breaking in analogy to electroweak symmetry breaking 
in the MSSM, provided the coupling strength is comparable to the top quark 
Yukawa coupling~\cite{Cvetic96}.},
\be
\label{SDD}
   S D \ovl{D},
\ee 
but they also allow new {\bf B} and {\bf L} violating (though {\bf B}--{\bf L} 
conserving) terms, 
\be
\label{QQD}
   q q D, \hspace{20pt} 
   \ovl{d}\ovl{D}\ovl{u}.
\ee
Together with the {\bf B} and {\bf L} conserving operators,
\be
   l q \ovl{D},         \hspace{20pt} 
   e^+ D \ovl{u},       \hspace{20pt} 
   \ovl{\nu} D \ovl{d},
\ee
dimension~4 proton decay is reintroduced through the exchange of $D$ and 
$\ovl{D}$ quarks. 

Clearly, one can avoid this problem by restricting oneself to $SO(10)$ and 
identifiying the \uonep with the $U(1)_\chi$ appearing in $SO(10) \rightarrow 
SU(5) \times U(1)_\chi$, which likewise forbids the terms in Eq.~(\ref{LLE}). 
The ordinary fermions and the $\ovl{\nu}$ complete anomaly free {\bf 16} 
representations (avoiding the extra singlet quarks), and one has to add a pair 
of Higgs doublets which by anomaly cancellation are required to have equal and 
opposite \uonep charges. As a result, a primordial $\mu$-term is allowed, and 
any term of the form (\ref{SHH}) can only include a \uonep neutral singlet 
field, $S$. Therefore, both aspects of the \uonep solution to the $\mu$-problem
(exclusion of an elementary $\mu$-term and generation of an effective
$\mu$-term at the TeV scale) are lost, while there is no problem with too rapid
proton decay. This is the reversed situation compared to the $U(1)_\psi$ case.

One possibility is to ignore the $\mu$-problem. Models containing an anomaly 
free \uonep symmetry designed to stabilize the proton have been discussed by 
various authors~\cite{Font89,Weinberg82,Chamseddine95}. Conversely, 
non-anomalous $U(1)^\prime$s addressing the $\mu$-problem without reference to 
proton decay are discussed in Ref.~\cite{Cvetic97A}. An interesting solution to
both problems is to flip the signs of the hypercharge assignments of the $D$ 
and $\ovl{D}$ quarks relative to the $E_6$ case~\cite{Aoki99}. This does not 
alter the SM anomaly conditions, and forbids the dangerous terms in 
Eq.~(\ref{QQD}). However, it implies the existence of stable fractionally 
charged baryons with TeV scale masses. Such states have not been observed, yet,
and lead to serious cosmological problems\footnote{I am indebted to Michael 
Pl\"umacher for discussions on this point.}. If the reheating temperature after
inflation is of the order of the mass of these exotic particles or above, their
hadronic interactions give rise to rather large number densities of these 
stable hadrons~\cite{Wolfram79}, orders of magnitude above experimental 
bounds~\cite{Smith82}. Lowering the reheating temperature does not solve 
the problem either, since successful baryogenesis probably requires a reheating
temperature after inflation at least of the order of the critical temperature 
of the electroweak phase transition of about 100~GeV~\cite{Riotto99}. Then 
a significant number density of TeV scale particles can still be produced by 
the high-energy tail of thermal particle distributions, or during reheating 
itself~\cite{Chung99}. The present work is therefore devoted to models without
fractionally charged states, i.e., $SU(5)$ charge quantization is imposed,
\be
   Q + \frac{{\cal C}_3}{3} = Q_Y + \frac{{\cal C}_2}{2} + \frac{{\cal C}_3}{3}
   \in \integer,
\label{quantem}
\ee
where $Q$ is electric charge, $Q_Y$ is hypercharge, and where ${\cal C}_2$ is 
an even (odd) integer for vector (spinor) representations of $SU(2)$. 
Similarly, ${\cal C}_3$ is the triality class of $SU(3)$ (e.g.\ $\pm 1$ for 
quarks and antiquarks).

\begin{table}[t]
\centering
\begin{tabular}{|ccrr|ccrr|}
\hline
$\left(\ba{c} \nu \\ e^-      \ea\right)$ & $-1$ &       &       &
$\ba{c} \ovl{\nu} \\ e^+      \ea$& $\ba{c} +5/3 \\ +1/3 \ea$    &
$\ba{r} +\tilde{a}\\-\tilde{a}\ea$& $\ba{r} +2 \tilde{b} \\ +\tilde{b}\ea$ \\
\hline
$\left(\ba{c}  u  \\  d      \ea\right)$  & +1/3 &       & $+\tilde{b}$  &
$\ba{c} \ovl{u}   \\ \ovl{d} \ea$ & $\ba{c} +1/3 \\ - 1  \ea$    &
$\ba{r} +\tilde{a}\\-\tilde{a}\ea$& $\ba{r} +\tilde{b}\\ {}\ea$   \\
\hline\hline
$\left(\ba{c}  N  \\ E^-     \ea\right)$  & +2/3 &$+\tilde{a}$&$-\tilde{b}$&
$\ba{c}        D  \\ \ovl{D} \ea$ & $\ba{c} -2/3 \\ +2/3 \ea$    &
$\ba{r}           \\         \ea$ & $\ba{r} -2\tilde{b}\\-\tilde{b}\ea$ \\
\hline
$\left(\ba{c} E^+ \\ \ovl{N} \ea\right)$  &$-2/3$& $-\tilde{a}$ &$-2\tilde{b}$&
$S$ & 0 & & $\ba{r} +3\tilde{b} \ea$ \\
\hline
\end{tabular}
\caption[]{$U(1)^\prime$ charges of a {\bf 27} representation for an arbitrary
$E_6$ boson~\cite{Erler00}. The upper half of the Table with $\tilde{b}=0$ 
corresponds to a family in $SO(10)$ GUT. The doublets on the left-hand side are
the fields $l$, $q$, $h_d$, and $h_u$, respectively. The special cases 
$(\tilde{a},\tilde{b}) = (0,0)$, $(0,-4/3)$, and $(-5/3,0)$ correspond,
respectively, to the $U(1)_\chi$, the $U(1)_\psi$, and the $U(1)_Y$ (up to 
normalization).}
\label{E6}
\end{table}

All of the $E_6$ inspired models discussed above have another unwanted feature
in the context of gauge coupling unification, which requires the addition of 
an extra pair of $SU(2)$ doublets. By virtue of anomaly cancellation these must
be non-chiral and one would therefore need a mechanism preventing them from 
receiving high scale tree-level masses. This is not a problem in the context of
string models which predict the absence of fundamental mass parameters in 
the low energy effective Lagrangian by conformal invariance. However, as these 
models stand, the extra doublets would then be massless. One could couple them 
to an overall singlet receiving a VEV, but in general this is not protected to 
be of electroweak size. Alternatively, one can assume that the extra doublets 
develop VEVs, as well, but it seems difficult to arrange that all chargino and
neutralino masses are compatible with the experimental lower mass limits as 
some masses would be predicted to be {\em strictly\/} of order the $Z$ boson 
mass, $\lsim M_Z$. This paper is therefore devoted to stabilize electroweak 
scale physics by the principles of gauge and supersymmetry, while assuring that
all fields receive large enough masses after symmetry breaking. 

An alternative way to remove the dangerous terms in Eq.~(\ref{QQD}) would be
to choose a particular linear combination of $U(1)_\chi$, $U(1)_\psi$, and 
$U(1)_Y$. Table~\ref{E6} shows that the condition,
\be
\label{ASO10}
   \tilde{a} + 2 \tilde{b} + {5\over 3} = 0,
\ee
allows a Dirac mass term, $D \ovl{d}$, so that $D$ and $\ovl{d}$ can be removed
from the spectrum. Baryon number is now conserved because the trilinear terms 
in Eq.~(\ref{LLE}) are still forbidden, and effects from the ones in 
Eq.~(\ref{QQD}) decouple. The role of $\ovl{d}$ is now played by $\ovl{D}$, but
since $l q \ovl{D}$ is the only type of operator involving it, $l$ must act as 
a Higgs superfield, and trade its role with $h_d$. However, Eq.~(\ref{ASO10}) 
also implies that $l$ and $h_u$ have equal and opposite \uonep charges and 
furthermore allows $\ovl{\nu}$ to decouple from the spectrum. Hence, we recover
the $SO(10)$ case (with $\ovl{\nu}$ replaced by $S$) and the $\mu$-problem 
which comes with it. Eq.~(\ref{ASO10}) defines therefore an alternative set of 
$SO(10)$ models within $E_6$, in analogy with the alternative left-right 
model~\cite{Ma87}.

\section{Anomaly cancellation in the presence of an extra \uonep}
\label{1singlet}
Consider the \uonep charge assignment in Table~\ref{mysol}. Indicated are three
generations of ordinary matter fields, one pair of Higgs doublets, $h_d$ and 
$h_u$, appropriate to break electroweak symmetry, one singlet Higgs, $S$, for 
\uonep symmetry breaking, and further singlet fields\footnote{I will refer to 
a singlet as a {\em right-handed neutrino}, $\ovl{\nu}$, if it admits 
the Yukawa coupling in Eq.~(\ref{Yukawa}).}, $T_i$, which are (initially) 
assumed not to receive VEVs. Shown are also examples of extra matter fields,
where $E^-$ are charged lepton singlets, $L$ are lepton doublets, $U$ and $D$ 
are up-type and down-type quark singlets, $Q$ are quark doublets, and where 
$E^+$, $\ovl{L}$, $\ovl{U}$, $\ovl{D}$, and $\ovl{Q}$ are their mirror 
partners. $W$ and $G$ are neutral fields under the $U(1)_Y$ transforming in 
the adjoint representations of $SU(2)$ and $SU(3)$, respectively. These are 
the fields which will play a role in the models surviving the analysis in 
Section~\ref{2singlets}, but initially any field respecting Eq.~(\ref{quantem})
is allowed. I assume family universality for the ordinary fermions and also 
the exotic fields, i.e., the new fields have identical \uonep charges in 
the cases of multiplicities greater than one, but this assumption is not 
crucial and can be relaxed easily. 

\begin{table}[t]
\centering
\begin{tabular}{|c|c|c||c|c|c|}
\hline
& $Q_Y$ & $Q^\prime$ & & $Q_Y$ & $Q^\prime$ \\ 
\hline\hline
$\left(\ba{c} \nu \\ e^-     \ea\right)$  & $-1/2$ &   $a$ &
$\ba{c} \ovl{\nu} \\ e^+     \ea$ & $\ba{c} 0 \\ +1 \ea$ &
$\ba{c}    - (a + c_2)  \\    - (a + c_1) \ea$         \\
$\left(\ba{c}  u  \\  d      \ea\right)$  & +1/6 &        $b$ &
$\ba{c} \ovl{u}   \\ \ovl{d} \ea$ & $\ba{c} -2/3 \\ +1/3 \ea$ &
$\ba{c}    - (b + c_2)  \\    - (b + c_1) \ea$                \\
\hline
$h_d$ &  $-1/2$  & $c_1$ &   $h_u$   &  $+1/2$  & $c_2$ \\
\hline\hline
 $D$  & $-1/3$ & $d_1$ & $\ovl{D}$ & $+1/3$ & $-(s+d_1)$ \\ 
 $L$  & $-1/2$ & $d_2$ & $\ovl{L}$ & $+1/2$ & $-(s+d_2)$ \\ 
 $Q$  & $+1/6$ & $d_3$ & $\ovl{Q}$ & $-1/6$ & $-(s+d_3)$ \\ 
 $U$  & $+2/3$ & $d_4$ & $\ovl{U}$ & $-2/3$ & $-(s+d_4)$ \\ 
$E^-$ &  $-1$  & $d_5$ &   $E^+$   &  $+1$  & $-(s+d_5)$ \\
 $W$  &    0   & $-s/2$&    $G$    &    0   & $-s/2$     \\
\hline
 $S$  &    0   &  $s$  &   $T_i$   &    0   & $t_i$ \\
\hline
\end{tabular}
\caption[]{$U(1)_Y$ and \uonep charge assignments for the MSSM matter fields 
including right-handed neutrinos and the two Higgs doublets (upper half), 
a number of singlets $S$ and $T_i$, and extra {\em exotic\/} fields. Three 
generations of ordinary fermions are assumed and generation indices suppressed.
Table~\ref{E6} corresponds to $a = - 1$ (normalization), 
$c_1 + c_2 = - s = - (a + 3b)$, and $d_1 = - 2 b$.}
\label{mysol}
\end{table}

As usual, all ordinary charged leptons and quarks are expected to become 
massive through the Yukawa couplings\footnote{While $m_u = 0$ is strongly 
disfavored, it is not firmly ruled out. I will comment on this case later in 
this Section.} in Eq.~(\ref{Yukawa}). On the other hand, the extra fields, such
as $D$ and $E$, are assumed to get vector-like (with respect to the SM group) 
masses induced by trilinear couplings with $S$ and their respective $SU(5)$ 
mirror partners, as for example in Eq.~(\ref{SDD}). This way their 
contributions to oblique parameters decouple. If an irreducible representation 
is real, ${\bf r} = {\bf \ovl{r}}$, and has $Q^\prime = - s/2$, it can acquire 
a supersymmetric Majorana mass through a trilinear coupling to $S$. Examples 
are the fields $W$ and $G$, and in this case it suffices to add a single copy. 
Finally, I demand,
\be
   s = - (c_1 + c_2) \neq 0,
\label{sneq0}
\ee
to guarantee a solution to both aspects of the $\mu$-problem. 

Since the exotic matter is vector-like with respect to the SM gauge group it 
does not spoil the anomaly cancellation present there. Triangle diagrams with 
two SM (i.e., either $SU(3)$, or $SU(2)$, or $U(1)_Y$) and one \uonep gauge 
fields as external legs yield three conditions which have to be satisfied. 
At first sight one would expect that there are many solutions given the many 
free parameters in Table~\ref{mysol}, but as it will turn out there are none. 
For example, if the exotic matter consists of $k_5$ copies of 
${\bf 5} + {\bf \ovl{5}}$ representations of $SU(5)$, the mixed $SU(3)$/\uonep 
anomaly condition reads,
\be
   (k_5 - 3) (c_1 + c_2) = 0,
\label{cond1}
\ee
and together with Eq.~(\ref{sneq0}) fixes $k_5 = 3$. Eq.~(\ref{cond1}) is
independent of the $d_i$, and so are the mixed $SU(2)$/\uonep and 
$U(1)_Y^2$/\uonep conditions, which are, respectively,
\begin{eqnarray}
             (k_5 +  1) (c_1 + c_2) + 3 (a + 3 b) = 0, 
\label{cond2} \\ 
        ({10\over 3} k_5 - 14) (c_1 + c_2) - 6 (a + 3 b) = 0. 
\label{cond3}
\end{eqnarray}
This is solved only by the $SO(10)$-type relation, $a = -3 b$, and for equal 
and opposite Higgs charges, $c_1 = - c_2$.

This conclusion cannot be avoided even if we generalize to arbitrary extra 
matter fields of the form ${\bf N} + {\bf \ovl{N}}$, where {\bf N} is any 
(in general reducible) representation of $SU(5)$. 
Eqs.~(\ref{cond1})--(\ref{cond3}) still apply, but $k_5$ is now defined as 
the {\em index\/} of ${\bf N}$,
\be
 k_5 = 2 \sum\limits_r \frac{n_r\, {\rm dim\,} {\bf r}} {{\rm dim\,} SU(5)} C_r
\label{index}
\ee
where the sum is over the irreducible representations, ${\bf r} \in {\bf N}$. 
$n_r$ are the multiplicities ($n_r = 1/2$ for Majorana types) and $C_r$ is 
the second-order Casimir invariant of representation {\bf r}. One can also 
allow extra matter fields in incomplete representations of $SU(5)$. However, 
preservation of gauge coupling unification enforces the {\em differences\/} of 
the one-loop $\beta$-function coefficients to be the same as in 
the MSSM\footnote{I will refer to such configurations as {\em quasi-complete\/}
$SU(5)$ representations.}. This leaves only one adjustable parameter, the index
$k_5$, and the anomaly conditions remain unchanged.

The conclusion, $c_1 + c_2 = 0$, is also independent of the assumption of 
family universality. If, for example, the \uonep charges of the third 
generation were different from the two lighter ones and associated with 
parameters $a^\prime$ and $b^\prime$ (cf.\ Table~\ref{mysol}), then this would 
merely result in the replacement,
\be
   3 (a + 3 b) \longrightarrow 2 a + a^\prime + 3 (2 b + b^\prime),
\ee
in Eqs.~(\ref{cond2}) and (\ref{cond3}).

A more interesting case~\cite{Langacker00} can be constructed from the one in 
the previous paragraph by replacing $-(b + c_2)$ by $-(b^\prime + c_2)$ for 
the quark singlets $\ovl{u}$ and $\ovl{c}$, and conversely for $\ovl{t}$. This 
results in a zero eigenvalue for the up-type quark mass matrix, and would 
either require $m_u = 0$, or that a non-vanishing $m_u$ can be generated 
radiatively, non-perturbatively, or by some other mechanism. Family 
non-universal \uonep charges can also induce significant flavor changing 
neutral current effects~\cite{Langacker00A}. If one allows for 
this possibility, Eqs.~(\ref{cond1})--(\ref{cond3}) are replaced by,
\be
   \left( \ba{ccc}
       0 &        + 1  &             k_5 -  3 \\
      +1 &        - 3  &             k_5 +  1 \\
      -2 & {34\over 3} & {10\over 3} k_5 - 14
   \ea \right) 
   \left( \ba{c} 2 a + a^\prime + 9 b \\ b - b^\prime \\ c_1 + c_2 \ea \right)
    = 0.
\label{anomaly1}
\ee
However, the determinant of this matrix is non-vanishing ($= - 4$) 
independently of $k_5$, and there is only the trivial solution implying 
$b = b^\prime$, i.e., nothing new. 

The setup discussed in this section does not seem to provide solutions to 
the $\mu$-problem. It should be stressed, however, that this conclusion depends
crucially on the additional assumptions of gauge coupling unification and 
chirality. For example, the model in Ref.~\cite{Aoki99} is chiral, but fails 
to produce unification; alternatively, if unification is enforced by adding 
an extra pair of doublets it turns non-chiral. Another example can be 
constructed with the assignments in Table~\ref{mysol}. Suppose the exotic 
matter consists of the fields $(Q+\ovl{Q}) + (U+\ovl{U})$. Eq.~(\ref{cond1}) is
then satisfied ($k_5 = 3$), and Eq.~(\ref{cond3}) becomes equivalent to 
Eq.~(\ref{cond2}), which are solved by, $ 4 (c_1 + c_2) = - 3 (a + 3 b)$.
The {\em quadratic\/} anomaly condition from triangle graphs with one 
hypercharge and two \uonep gauge bosons in the external legs depends on $d_3$ 
and $d_4$ and can easily be solved. Extra singlets, $T_i$, can be added 
to satisfy the {\em cubic\/} (pure $U(1)^\prime$) and {\em trace\/} (mixed 
gravitational/$U(1)^\prime$) anomaly conditions. This model is chiral but 
the gauge couplings do not unify. Unification can be arranged by adding 
an $E^\pm$ pair to complete a ${\bf 10} + {\bf \ovl{10}}$ representation of 
$SU(5)$, but it would have to carry equal and opposite \uonep charges, spoiling
the chirality of the model. Thus, this model mirrors the situation in 
Ref.~\cite{Aoki99} except that it has no fractionally charged baryons.

\section{Two Higgs singlet solutions}
\label{2singlets}
Suppose now that besides $S$ another MSSM singlet field, $T = T_0$, with 
$Q^\prime = t = t_0$, develops a VEV. A subset of the exotic matter fields, 
${\bf M} + {\bf \ovl{M}} \subset {\bf N} + {\bf \ovl{N}}$, could then acquire 
TeV scale masses through trilinear couplings to $T$, provided the \uonep 
charge assignments of the fields in ${\bf M}$  and ${\bf \ovl{M}}$ are 
changed\footnote{In general, this explicitly violates the assumption of 
universality w.r.t.\ the exotic fields which was made in 
Section~\ref{1singlet}.} from $d_i$ and $ - (s + d_i)$ to $e_i$ and 
$ - (t + e_i)$, respectively. The anomaly conditions in 
Eqs.~(\ref{cond1})--(\ref{cond3}) now change to,
\be
   \left( \ba{ccc}
       0 &              k_5 - k_3  -  3 &             k_3 \\
      -1 &              k_5 - k_2  +  1 &             k_2 \\
      +2 & {10\over 3} (k_5 - k_1) - 14 & {10\over 3} k_1
   \ea \right) 
   \left( \ba{c} 3 (a + 3 b) \\ s \\ t \ea \right) = 0,
\label{anomaly}
\ee
where $k_3$ is the $SU(3)$ index of ${\bf M}$,
$k_3\, {\rm dim\,} SU(3) = 2 \sum_r m_r\, {\rm dim\,}{\bf r\,} C_r$. Similarly,
$k_2$ is the $SU(2)$ index of ${\bf M}$, and $k_1 = 6/5\, \Tr_{\bf M} Q_Y^2$. 
The $SU(5)$ charge quantization condition~(\ref{quantem}) implies,
\be
   {5\over 6} k_1 - {1\over 2} k_2 - {1\over 3} k_3 \equiv k_0 \in \integer.
\label{k0}
\ee
Notice, that $k_3$, $k_2$, and $k_0$ are integers, and that by definition, 
$0 \leq k_j \leq k_5$ for $1 \leq j \leq 3$, which implies the bound 
$|k_0| \leq 5 k_5/6$. If the determinant of the matrix in Eq.~(\ref{anomaly}) 
is non-vanishing, there is only the unwanted solution, implying 
$c_1 + c_2 = 0$. The matrix becomes singular for,
\be
k_3 = (k_5 - 3) (k_0 + k_2 - k_3).
\label{det}
\ee
Using the definitions,
\be
  \sigma \equiv {(a + 3 b)\over s}, \hspace{20pt}  \tau \equiv {t\over s}, 
\ee
the solutions can be classified as follows:
\begin{itemize} 
\item The case $k_3 = k_0 + k_2 = 0$ implies $5 k_1 = - 3 k_2$, and therefore
      $k_0 = k_1 = k_2 = \sigma = \tau = 0$ independently of $k_5$, which will 
      be rejected because it corresponds essentially to the non-chiral
      situation in Section~\ref{1singlet}. This is the only possibility for 
      $k_5 = 1$.
\item If $k_5 \geq 5$, Eq.~(\ref{det}) implies that $k_3$ is an integer 
      multiple of $k_5 - 3$, and $k_j \leq k_5$ requires $k_3 = k_5 - 3$ and 
      $k_0 + k_2 = k_5 - 2$. One has $3 \sigma = (k_5 - k_2 + 1)$ and 
      $\tau = 0$, i.e., these models are also non-chiral since the fields 
      within the ${\bf M} + {\bf\ovl{M}}$ representation are not tied to \uonep
      symmetry breaking. Such solutions exist for $k_5 = 3$ (e.g., the three 
      generation $E_6$ model with an extra pair of doublets) and $k_5 = 4$, as 
      well, but they will be discarded.
\item $k_5 = 2$ implies $k_0 + k_2 = 0$, $k_3 > k_2$, $3 \sigma = 3 + k_2/k_3$ 
      and $\tau = 1 + 1/k_3$. There are three solutions, but the case 
      $(k_3,k_2,k_1,k_0) = (2,1,1/5,-1)$ cannot be made consistent with 
      the $SU(5)$ quantization condition.
\item $k_5 = 3$ implies $k_3 = 0$, $3 \sigma = 3 + k_0/(k_0 + k_2)$ and 
      $\tau = 1 - 1/(k_0 + k_2)$. There are eight solutions of this type. 
\item $k_5 = 4$ implies $k_3 = 2$, $k_0 + k_2 = 4$, $\sigma = 5/3 - k_2/6$ and 
      $\tau = 1/2$. There are two solutions, but the case 
      $(k_3,k_2,k_1,k_0) = (2,3,19/5,1)$ cannot be made consistent with 
      the $SU(5)$ quantization condition.
\end{itemize}
Thus, there are an infinite number of non-chiral solutions, but only 11~chiral 
ones, which are collected in Table~\ref{11cases}. All of them predict 
$\tau = t/s \geq 0$, implying that there are no $D$-flat scalar field 
directions. Furthermore, all of them predict $k_5 \leq 4$, which guarantees 
that the gauge couplings remain perturbative up to the unification scale, while
$k_5 = 5$ would yield non-perturbative unification~\cite{Hempfling95}.

\begin{table}[t]
\centering
\begin{tabular}{|c|c|c|c|c|r|c|c|r|c|c|}
\hline
solution & $k_5$ & $k_3$ & $k_2$ & $k_1$ & $k_0$ & $\sigma$ & $\tau$ &
Yukawa couplings to $T$ & $\omega_1$ & $\omega_3^{\rm max}$ \\
 \hline
   I & 2 & 1 & 0 &  2/5 & 0 & 1 &  2  &        $ (D+\ovl{D})$ &  0  & $-3/2$ \\
  II & 2 & 2 & 0 &  4/5 & 0 & 1 & 3/2 &        $2(D+\ovl{D})$ & 1/2 &  23/16 \\
 III & 3 & 0 & 2 &  6/5 & 0 & 1 & 1/2 &        $2(L+\ovl{L})$ & 3/2 &   3/2  \\
  IV & 3 & 0 & 3 &  9/5 & 0 & 1 & 2/3 &        $3(L+\ovl{L})$ & 4/3 &  31/27 \\
   V & 4 & 2 & 4 & 16/5 & 0 & 1 & 1/2 &all but $2(D+\ovl{D})$ & 3/2 &   9/16 \\
\hline
$E_6$& 3 & 0 & 1 &  3/5 & 0 & 1 &  0  &        $ (L+\ovl{L})$ &  2  &    2   \\
\hline
\cline{1-11}
  VI & 3 & 0 & 3 &  3/5 & $-1$ &  5/6  & 1/2 & $       W      + (L+\ovl{L})$ \\
 VII & 3 & 0 & 1 &  9/5 &   1  &  7/6  & 1/2 & $  (L+\ovl{L}) +   E^\pm$     \\
VIII & 3 & 0 & 2 & 12/5 &   1  & 10/9  & 2/3 & $2 (L+\ovl{L}) +   E^\pm$     \\
  IX & 3 & 0 & 3 &   3  &   1  & 13/12 & 3/4 & $3 (L+\ovl{L}) +   E^\pm$     \\
   X & 3 & 0 & 0 & 12/5 &   2  &  4/3  & 1/2 & $                2 E^\pm$     \\
  XI & 3 & 0 & 1 &   3  &   2  & 11/9  & 2/3 & $  (L+\ovl{L}) + 2 E^\pm$     \\
\cline{1-9}
\end{tabular}
\caption{Classification of solutions to the conditions of anomaly cancellation,
chirality, $SU(5)$ charge quantization, and gauge coupling unification. It is 
assumed that two fields participate in \uonep gauge symmetry breaking. For 
comparison, the non-chiral three generation $E_6$ solution with an extra 
neutral ($t = 0$) Higgs singlet giving mass to an extra vector-like pair of 
doublets ($k_2 = 1$) is also shown. $\omega_1$ and $\omega_3$ are shown for 
${\bf 5} + {\bf \ovl{5}}$ representations only, in which case they are defined 
in Eqs.~(\ref{trace}) and (\ref{cubic}), respectively. The column before these
specifies the fields receiving masses through trilinear couplings to $T$ (cf.\ 
Table~\ref{exotic} below). In addition, $2(L+\ovl{L})$ fields can be replaced 
by the combination $W + E^\pm$, where applicable.}
\label{11cases}
\end{table}

In general, each solution in Table~\ref{11cases} can be realized with various
exotic matter contents. Observe that the configurations of $k_i$ in the Table
restrict the possible matter content to the fields displayed in 
Table~\ref{mysol}. These fields can be combined into quasi-complete $SU(5)$ 
representations shown in Table~\ref{exotic}.

\begin{table}[t]
\centering
\begin{tabular}{|c|c|c|c|c|c|c|c|c|c|}
\hline
representation & $k_5$ & $D+\ovl{D}$ & $L+\ovl{L}$ & $Q+\ovl{Q}$ & $U+\ovl{U}$ 
& $E^\pm$ & $W$ & $G$ & solution(s) \\
\hline
 1 & 2 & 1 &---&---& 1 &---& 1 &---& I                \\
 2 & 2 & 2 &---&---&---& 1 & 1 &---& I, II            \\
 3 & 2 & 2 & 2 &---&---&---&---&---& I, II            \\
\hline
 4 & 3 &---& 1 &---&---& 2 & 1 & 1 & III, IV,  VI--XI \\
 5 & 3 &---& 3 &---&---& 1 &---& 1 & III, IV, VII--IX \\
 6 & 3 & 1 &---& 1 &---& 2 &---&---& X                \\
 7 & 3 & 2 & 1 &---& 1 &---& 1 &---& VI               \\
 8 & 3 & 3 & 1 &---&---& 1 & 1 &---& III, IV, VI, VII \\
 9 & 3 & 3 & 3 &---&---&---&---&---& III, IV          \\
\hline
10 & 4 & 2 &---&---& 2 &---& 2 &---& V                \\
11 & 4 & 2 & 1 & 1 &---& 2 &---&---& V                \\
12 & 4 & 3 &---&---& 1 & 1 & 2 &---& V                \\
13 & 4 & 3 & 2 &---& 1 &---& 1 &---& V                \\
14 & 4 & 4 &---&---&---& 2 & 2 &---& V                \\
15 & 4 & 4 & 2 &---&---& 1 & 1 &---& V                \\
16 & 4 & 4 & 4 &---&---&---&---&---& V                \\
\hline
\end{tabular}
\caption[]{Quasi-complete $SU(5)$ representations (preserving gauge coupling 
unification) which can be used to realize the solutions in Table~\ref{11cases}.
There are other configurations with $k_5 = 4$, but solution~V requires at least
two pairs of $D+\ovl{D}$ quarks. Similarly, the ${\bf 10} + {\bf \ovl{10}}$ 
representation of $SU(5)$ ($k_5 = 3$) cannot be used for any of the solutions 
in Table~\ref{11cases} and has been omitted.}
\label{exotic}
\end{table}

In the remainder of this Section and in the next Section I will focus on exotic
matter in complete ${\bf 5} + {\bf \ovl{5}}$ representations of $SU(5)$, unless
noted otherwise. In this case one can write the sum over the other singlet 
charges, which is fixed by the trace condition, $\Tr\; Q^\prime = 0$, as
\be
   \omega_1 \equiv {1\over s} \sum\limits_{i=1} t_i = 
   5 k_5 + (\tau - 1)(3 k_3 + 2 k_2 - 1) - 12.
\label{trace}
\ee
Interestingly, when 3 right-handed neutrinos are included, $\omega_1$ is
completely determined within each model. For ${\bf 5} + {\bf \ovl{5}}$ 
representations one has $k_0 = 0$, and then Eqs.~(\ref{sneq0}), 
(\ref{anomaly}), and~(\ref{k0}) imply the relation,
\be
   s = - (c_1 + c_2) = (a + 3 b),
\label{5plus5}
\ee
which is familiar from the $E_6$ assignment in Table~\ref{E6}. Thus, most of 
the phenomenological studies of $E_6$ inspired $Z^\prime$ bosons apply here,
and also to the other solutions with $k_0 = 0$. To simplify the algebra, I 
impose the orthogonality condition, $\Tr\; Q_Y Q^\prime = 0$, which can be 
relaxed at a later stage. Combined with the anomaly condition quadratic in 
the \uonep charges\footnote{$\Tr\; Q_Y {Q^\prime}^2$ also vanishes if 
$s = 0$.}, $\Tr\; Q_Y {Q^\prime}^2 = 0$, I find,
\bea \hspace{-20pt}
   2 (k_3 e_1 + k_2 e_2) &=& 9 \frac{(a - b) + (c_1 - c_2)}{\tau - 1} - 
   \tau (k_3 + k_2) s, 
\label{econd} \\ \hspace{-20pt}
   2 [(k_3 - k_5) d_1 + (k_2 - k_5) d_2] &=& \frac{(6\tau + 3)(a - b) +
   (7\tau + 2)(c_1 - c_2)}{\tau - 1} + [2 k_5 - (k_3 + k_2)] s.
\label{dcond}
\eea
Similarly, the cubic condition, $\Tr\; {Q^\prime}^3 = 0$, reads, 
\be
   \omega_3 \equiv {1\over s^3} \sum\limits_{i=1} t_i^3 = 
   5 k_5 + (\tau^3 - 1)(3 k_3 + 2 k_2 - 1) - {3\over 4} \left[ 3 + 9 \left( 
   {a - b\over s} \right)^2 + 4 \left( {c_1 - c_2\over s} \right)^2 \right] +
\label{cubic}
\ee
$$ 
3 \left[ 3  \tau  k_3  {e_1\over s} \left( {e_1\over s} + \tau \right) 
       + 2  \tau  k_2  {e_2\over s} \left( {e_2\over s} + \tau \right) 
       + 3 (k_5 - k_3) {d_1\over s} \left( {d_1\over s} +    1 \right)
       + 2 (k_5 - k_2) {d_2\over s} \left( {d_2\over s} +    1 \right)\right] ,
$$
and one can show that,
\be
  \omega_3 \leq 
  {1\over 4} \left[ 5 (k_5 - 1) + (\tau^3 - 1)(3 k_3 + 2 k_2) \right] - \tau^3.
\label{cubic_max}
\ee
The inequality~(\ref{cubic_max}) is saturated when,
\be
    a  =  b  =   {s\over 4}, \hspace{50pt}  
   c_1 = c_2 = d_1 = d_2 = - {s\over 2}, \hspace{50pt}  
   e_1 = e_2 = - {\tau\over 2} s,
\label{zpsi}
\ee
which yields a class of $Z^\prime$ bosons which couple like the $Z_\psi$ to 
the ordinary fermions and the Higgs doublets. This is encouraging as the recent
analysis~\cite{Erler00} of precision data revealed an excellent fit to 
the $Z_\psi$ assuming some amount of mixing with hypercharge. Recall that 
\uonep subgroups from $E_6$ are frequently discussed because they represent 
manifestly anomaly free examples (and because they arise in popular GUT and 
string models), but the $E_6$ gauge symmetry and Yukawa coupling relations are 
usually explicitly broken to avoid serious conflicts with observation (such as 
the proton lifetime). Here we recover $U(1)^\prime \equiv U(1)_{\tilde\psi}$ 
symmetries which are very similar to the $U(1)_\psi$ from a very different 
bottom-up approach. 

Relaxing the orthogonality condition, $\Tr\; Q_Y Q^\prime = 0$, and using 
the trivial solution to Eq.~(\ref{anomaly}), $a + 3 b = c_1 + c_2 = s = t = 0$,
one can construct another family of special solutions, $U(1)_{\tilde\chi}$, 
where the MSSM fields have charges as in the $U(1)_\chi$ case. While they would
yield non-chiral pairs of Higgs doublets, these solutions could be interesting 
if they enter linear combinations with the $U(1)_{\tilde\psi}$ or if 
the corresponding $Z_{\tilde\chi}$ coexists with the $Z_{\tilde\psi}$. 
The quadratic anomaly condition now reduces to
\be
(k_3 - k_5) d_1 + (k_2 - k_5) d_2 - \tau (k_3 e_1 + k_2 e_2) + 2 (a + c_1) = 0,
\label{chicond}
\ee
where solutions with $c_1 = a$ correspond to hypercharge and $c_1 = - 2 a/3$ to
the $U(1)_{\tilde\chi}$. Note, that there are other solutions to the anomaly 
constraints with $s = 0$, but only those satisfying Eq.~(\ref{chicond}) can mix
with the chiral $U(1)^\prime$s.

The encounter of the $U(1)_{\tilde\psi}$ and $U(1)_{\tilde\chi}$ groups is also
encouraging from a top-down perspective. There are classes of string models 
where gauge coupling unification is predicted without the appearance of a full 
GUT group in the zero-slope limit. Examples are heterotic string constructions 
where the gauge group factors considered are realized at the same Kac-Moody 
level, $k$, or open string models with the group factors arising from the same 
$D$-brane. The $E_8 \times E_8$ heterotic string is particularly suited here, 
because it offers the prerequisite $E_6$ subgroups. The reference $E_8$ 
symmetry implies the relevant matching (unification) conditions, as well as 
quantization conditions for its $U(1)$ subgroups. The latter are given by
\be
   Q_\chi - {6\over 5} Q_Y    \in \integer, \hspace{50pt}
   Q_\psi - {5\over 4} Q_\chi \in \integer, \hspace{50pt}
   Q_\phi - {4\over 3} Q_\psi \in \integer,
\label{quant}
\ee
where the last relation refers to the $U(1)_\phi$ defined by
$E_7 \rightarrow E_6 \times U(1)_\phi$. The normalizations are chosen such that
$Q_Y = 1$ for $e^+$ ($a = - 1/2$), $Q_\chi = 1$ for $\ovl{\nu}$ ($a = - 3/5$), 
$Q_\psi = 1$ for $S$ ($s = 1$), and $Q_\phi = 1$ for one of the $E_6$ singlets 
within the {\bf 56} of $E_7$. Eq.~(\ref{chicond}) implies for 
the $U(1)_{\tilde\chi}$ of solution~II, $2 d_2 + 3 e_1 + 2/5 = 0$, or when 
combined with the first Eq.~(\ref{quant}), $e_1 + 2/5 \in 2 \integer$. With 
the exception of two fields, the second Eq.~(\ref{quant}) is also satisfied.
But the $D$ quarks have the sign of their $U(1)_{\tilde\psi}$ charges reversed
and the $T$ field ($Q_{\tilde\chi} = 0$) has half-integer instead of integer 
$U(1)_{\tilde\psi}$ charge. It is not clear at present\footnote{In Kac-Moody 
level~1 models $SU(5)$ charge quantization can {\em never\/} be obtained 
unless a complete $SU(5)$ GUT group survives~\cite{Schellekens90}. On the other
hand in the level~2 models or Ref.~\cite{Erler96} the third Eq.~(\ref{quant}) 
is respected by all fields.} whether $E_8$ charge quantization, i.e., 
Eqs.~(\ref{quant}), are predicted in string models with gauge coupling 
unification and $SU(5)$ charge quantization. If so one would have to impose 
the condition $\tau \in \integer$, and only solution~I would survive. 

The $E_8$ embedding also allows one to fix the overall normalization (coupling 
strength) of the \uonep. In units in which an $E_8$ root vector $P$ has length 
$P^2 = 1$, one has $P^2 = 3/5$ for a {\bf 10} of $SU(5)$, $P^2 = 5/8$ for 
a {\bf 16} of $SO(10)$, $P^2 = 2/3$ for a {\bf 27} of $E_6$, and $P^2 = 3/4$ 
for a {\bf 56} of $E_7$. The {\bf 10} of $SU(5)$ is relevant here, because it 
contains the $SU(3) \times SU(2)$ singlet with unit charge (the $e^+$) and can 
therefore serve to properly normalize the hypercharge gauge coupling, $g_Y$, in
terms of one of the non-Abelian gauge couplings, $g$. Assuming 
the normalization in Eq.~(\ref{quant}) one finds the well-known result, 
$g_Y = \sqrt{3/5}\, g$. Notice, that this argument needs no reference to 
a complete GUT group like $SU(5)$, but it applies only to string constructions 
in which the embedding into $E_8$ is still traceable. Referring again to 
the normalization fixed by Eq.~(\ref{quant}) one finds in a similar way, 
$g_\chi = \sqrt{5/8}\, g$ and $g_\psi = \sqrt{2/3}\, g$. 

The reference $E_8$ root vector also allows to set lower limits on the possible
Kac-Moody levels, $k$. For example, the singlet $T$ of solution~II has 
$Q_{\tilde\psi} = 3s/2$, which implies $P_T^2 = 3/2 > 1$ and means that $T$ 
cannot be obtained in the massless sector if $k = 1$, while $k \geq 2$ would be
allowed by this consideration. Interestingly, the theorem about fractionally 
charged states in level $k = 1$ models~\cite{Schellekens90} implies the same 
conclusion.

\section{Phenomenological aspects}
\label{pheno}
The one-loop $\beta$-function coefficient for the \uonep is given by,
\be
   {\beta_1^\prime \over s^2} = 
   \omega_2 + 5 k_5 + (\tau^2 - 1)(3 k_3 + 2 k_2 + 1) + 6 +
   9 \left({a - b\over s} \right)^2 + 7 \left( {c_1 - c_2\over s} \right)^2 +
\label{quadr}
\ee
$$ 
2 \left[ 3        k_3  {e_1\over s} \left( {e_1\over s} + \tau \right) 
       + 2        k_2  {e_2\over s} \left( {e_2\over s} + \tau \right) 
       + 3 (k_5 - k_3) {d_1\over s} \left( {d_1\over s} +    1 \right)
       + 2 (k_5 - k_2) {d_2\over s} \left( {d_2\over s} +    1 \right)\right] ,
$$
where $\omega_2 = s^{-2} \sum_{t=i} t_i^2$. $\beta_1^\prime$ is minimized when 
Eqs.~(\ref{zpsi}) are satisfied, i.e., for the $U(1)_{\tilde\psi}$, in which 
case one finds,
\be
   \beta_1^\prime = {1\over 3} \left[ 5 (k_5 + 2) + (\tau^2 - 1)(3 k_3 + 2 k_2)
   + 2 (\tau^2 + \omega_2) \right] .
\label{quadr_max}
\ee
In Eq.~(\ref{quadr_max}) I have chosen the $E_6$ inspired normalization, 
$s^2 = 2/3$. For comparison, the $SU(3)\times SU(2)\times U(1)_Y$ one-loop
$\beta$-function coefficients are $\beta_3 = k_5 - 3$, $\beta_2 = k_5 + 1$, and
$\beta_1 = k_5 + 33/5$ (using $SU(5)$ normalization, $a^2 = 3/20$), 
respectively. 

For concreteness, I now focus on the $Z_{\tilde\psi}$ of solution~II which is 
reminiscent of an $E_6$ model with three families of {\bf 27}, but with one 
copy of $D + \ovl{D}$ quarks removed and the \uonep charges of the other two
copies altered. The trace and cubic anomaly conditions, $\omega_1 = 1/2$ and 
$\omega_3 = \omega_3^{\rm max} = 23/16$, can be satisfied by adding, for 
example, $SU(5)$ singlets with \uonep charges, $t_1 = t_2 = s$, $t_3 = 3s/4$, 
$t_{2 i + 2} = - s/2$, and $t_{2 i + 3} = s/4$, where $1 \leq i \leq 9$, or 
in short-hand notation, $t_i \sim [1^2,3/4,-(1/2)^9,(1/4)^9]$. There are no 
singlets with charges $-1$ or $-3/2$ in this choice, assuring that the crucial 
fields, $S$ and $T$, cannot acquire high scale masses. The quantum numbers for 
this case, $Z_{\tilde\psi}^{({\rm II})}$, are summarized in Table~\ref{psi2}. 

\begin{table}[t]
\centering
\begin{tabular}{|c|c|c||c|c|c|}
\hline
&$Q_Y$&$Q_{\tilde\psi}^{({\rm II})}$&&$Q_Y$&$Q_{\tilde\psi}^{({\rm II})}$ \\ 
\hline\hline
$\left(\ba{c} \nu \\ e^-     \ea\right)$  & $-1/2$ &  +1/4    &
$\ba{c} \ovl{\nu} \\ e^+     \ea$ & $\ba{c}   0  \\   +1 \ea$ &
$\ba{c}    +1/4   \\   +1/4  \ea$ \\
$\left(\ba{c}  u  \\  d      \ea\right)$  & +1/6 &  +1/4      & 
$\ba{c} \ovl{u}   \\ \ovl{d} \ea$ & $\ba{c} -2/3 \\ +1/3 \ea$ &
$\ba{c}    +1/4   \\   +1/4  \ea$ \\ 
\hline
$h_d$ & $-1/2$ & $-1/2$ & $h_u$ & +1/2 & $-1/2$ \\
\hline\hline
$\ba{c} D_{1,2} \\ L_{1,2} \\   S  \\ T_{1,2}^{+1} \\ T_{2i+2}^{-1/2} \ea$ &
$\ba{c}   -1/3  \\   -1/2  \\   0  \\    0    \\     0    \ea$ &
$\ba{c}   -3/4  \\   -1/2  \\  +1  \\   +1    \\   -1/2   \ea$ &
$\ba{c}\ovl{D}_{1,2}\\ \ovl{L}_{1,2}\\ T \\ T_3^{+3/4} \\ T_{2i+3}^{+1/4}\ea$ &
$\ba{c}   +1/3  \\   +1/2  \\   0  \\    0    \\     0    \ea$ &
$\ba{c}   -3/4  \\   -1/2  \\ +3/2 \\  +3/4   \\   +1/4   \ea$ \\
\hline
\end{tabular}
\caption[]{Charge assignment for the $Z_{\tilde\psi}$ of solution~II realized
with a ${\bf 5} + {\bf \ovl{5}}$ of $SU(5)$. The \uonep normalization
corresponds to $s = 1$.}
\label{psi2}
\end{table}

There are a total of 12 fields with quantum numbers like the right-handed 
neutrinos. Since these fields are protected from acquiring high scale Majorana 
masses, one predicts a Dirac mass matrix with three electroweak scale 
eigenvalues. I.e., there are unacceptably large neutrino masses, unless
the Yukawa couplings are chosen zero or tiny. A clean way to cure this problem
is to add the set of singlets\footnote{By virtue of the identity,
$n^3 - 4 (n - 1)^3 + 6 (n - 2)^3 - 4 (n - 3)^3 + (n - 4)^3 = 0$, such 
a configuration does not alter $\omega_1$ or $\omega_3$.}, 
$t_i \sim 3 [1,-(3/4)^4,(1/2)^6,-(1/4)^4]$. Now the fields with $t_i = \pm s/4$
(and some others) can form bilinear Dirac mass terms, resulting in a neutrino 
mass matrix with three zero eigenvalues, which correspond predominantly to 
the left-handed neutrinos provided the bilinear masses are much larger than 
the Yukawa mass terms. Clearly, the bilinear masses can be taken to infinity 
and the corresponding fields can be smoothly removed from the spectrum, 
yielding the residual {\em chiral\/} set of singlets (now including $S$, $T$, 
and the three $\ovl{\nu}$), $\sim [3/2,1^6,-(3/4)^{11},(1/2)^9]$.

There is another interesting solution which can be obtained by adding 
the fields, $t_i \sim 3 [-(5/4),1^4,-(3/4)^6,(1/2)^4,-(1/4)]$ to the ones in 
Table~\ref{psi2}. This results in the {\em chiral\/} singlet set, 
$\sim [3/2,-(5/4)^3,1^{15},-(3/4)^{17},(1/2)^3,(1/4)^9]$, where the fields with
$Q^\prime = - 5 s/4$ allow trilinear couplings with $S$ and the $\ovl{\nu}$. 
This once again results in three zero eigenvalues, and if the masses generated 
by $S$ are an order of magnitude larger than the ones from the standard Yukawa 
terms, these also correspond predominantly to the left-handed neutrinos, but 
the mixing with the massive states could induce a small amount of missing 
invisible $Z$ width, $\Gamma_{\rm inv}$. Indeed, the LEP 
Collaborations~\cite{Mnich99} currently report a shortage of about~0.5\% in 
$\Gamma_{\rm inv}$. Furthermore, the Fermi constant, $G_F$, extracted from 
$\mu$ decays could increase, significantly increasing the extracted Higgs boson
mass from electroweak fits (the central value of which is currently below 
the lower search limit~\cite{Erler99A}) but potentially deteriorating 
the quality of the global fit to all data. There can be other effects 
associated with this scenario, such as a violation of weak charged current 
universality. There are strong limits on the non-universality between the first
two families, however, driven by the well measured branching ratio, 
$\frac{{\cal B}(\pi\rightarrow \mu\nu_\mu)}{{\cal B}(\pi\rightarrow e\nu_e)}$.
Thus the effect in $\Gamma_{\rm inv}$ should be expected to be dominated by
the $\nu_\tau$ sector, in agreement with the experience that the third family 
Yukawa couplings are typically the largest. 

Yet another solution to the problem of neutrino mass would be to implement 
the see-saw mechanism which can be achieved by choosing a linear combination 
of the $U(1)_{\tilde\psi}$ and the $U(1)_{\tilde\chi}$, such that 
the $\ovl{\nu}$ are \uonep neutral and allowed to receive high scale Majorana
masses. As is well known, this also produces small but non-vanishing 
Majorana masses for the left-handed neutrinos, which is not the case for 
the scenarios in the previous two paragraphs. However, one should generally
expect non-vanishing masses for the left-handed neutrinos originating from
non-renormalizable terms suppressed by powers of the high mass 
scale~\cite{Langacker98}. Dimension~5 operators would yield the same kind of 
suppression as in the see-saw mechanism, but higher order suppressions (by 
intermediate scales) are conceivable\footnote{Employing the $Z^\prime$ scale, 
$M_{Z^\prime} \approx 800$~GeV~\cite{Erler00}, and the unification scale, 
$M \sim 2\times 10^{16}$~GeV, suggests neutrino masses of order 
$M_{Z^\prime}^2/M \sim 0.03$~eV, in excellent agreement with the mass scale
inferred from atmospheric neutrino oscillations~\cite{Fukuda98}. One might 
wonder whether this agreement not only signals the presence of a new physics 
scale close to the unification scale, but possibly even suggests the absence of
further scales below that.}.

As for baryon number violating interactions, the $U(1)_{\tilde\psi}$ symmetry 
of solution~II forbids the terms in Eq.~(\ref{QQD}), thus assuring proton 
stability at the renormalizable level. Dimension~5 operators can also lead to 
proton decay at rates conflicting with the experimental limits. However, 
the $D$-terms,
\be
        q       q  \ovl{d}^\ast, \hspace{20pt} 
        q       q  \ovl{D}^\ast, \hspace{20pt} 
        D       D  \ovl{u}^\ast,\hspace{20pt} 
   \ovl{u} \ovl{d}       D^\ast, \hspace{20pt} 
   \ovl{u} \ovl{D}       D^\ast, 
\label{QQDstar}
\ee
are also absent by virtue of the $U(1)_{\tilde\psi}$, and likewise for 
the $F$-terms, 
\be
      l    q q q, \hspace{20pt} 
     H_d   q q q, \hspace{20pt} 
      L    q q q, \hspace{20pt} 
     H_u   q D D, \hspace{20pt} 
   \ovl{L} q D D, \hspace{20pt} 
     e^+   D D D. \hspace{20pt} 
     e^+ \ovl{u} \ovl{u} \ovl{d}.
\ee
On the other hand, the $F$-terms,
\be
   \ovl{\nu}       q       q       D,  \hspace{20pt} 
   \ovl{\nu}  \ovl{u} \ovl{d} \ovl{D}, \hspace{20pt}
   T_i^{-3/4} \ovl{u} \ovl{d} \ovl{d}, \hspace{20pt}
   T_i^{+5/4} \ovl{u} \ovl{D} \ovl{D},
\label{QQDT}
\ee
are guaranteed to be absent only when there are no MSSM singlets with 
$Q^\prime = 5 s/4$, $s/4$, or $-3 s/4$. While this can be arranged, there is 
another dimension~5 $F$-term,
\be
   e^+ \ovl{u} \ovl{u} \ovl{D}, 
\label{DUUE}
\ee
which is similar to the operator $e^+ \ovl{u} \ovl{u} \ovl{d}$, present in 
the $R$-parity {\em conserving\/} version of the MSSM. There one has to assume
that its corresponding coupling strength is small, and taking into account 
suppressions from first generation Yukawa couplings, Cabibbo mixing, and phase
space, the resulting proton lifetime can be acceptable. In comparison, 
the term~(\ref{DUUE}) is less concerning, because at the renormalizable level
the $D$ quarks couple only (ignoring the possibility of 
a $T_i^{+1/2} \ovl{d} D$-term) via gauge interactions and the $TD\ovl{D}$-term,
both conserving ``$D$-number''. Therefore, observable proton decay is not 
expected, unless a large $\ovl{d}\ovl{D}$ mixing effect can be generated. 
Alternatively, the critical dimension~5 $F$-term in Eq.~(\ref{DUUE}) can be 
removed if there is an admixture of the $U(1)_{\tilde\chi}$. Thus, choosing 
$k_3 = k_5$ (or more generally coupling all of the exotic quarks to singlets 
other than $S$) is an efficient strategy to achieve a realistic proton 
lifetime.

\section{Conclusions}
\label{con}
I have presented a class of models with an additional non-anomalous \uonep in 
which gauge couplings unify, the $\mu$-problem can be solved, and electric 
charge is quantized. These models are chiral, protecting a large desert between
the electroweak (SUSY breaking) and unification (string) scales. Moreover, in
some of these models the \uonep is sufficient to avoid conflicts with 
experimental bounds on the proton lifetime. Within a deliberately traditional 
and conservative framework the analysis presented is rather general, but there 
are various directions which can be chosen to go beyond it.

For example, I assumed that two MSSM singlet fields participate in \uonep 
symmetry breaking. Only a finite number of solutions exist since the anomaly 
matrix in Eq.~(\ref{anomaly}) has to be singular. The analysis generalizes 
straightforwardly if one allows three (or more) singlets to acquire VEVs. In
this case an infinite number of solutions exist, a large (but finite) number
of which being consistent with {\em perturbative\/} gauge coupling unification.
A preliminary study of this class reveals that unlike the solutions in 
Table~\ref{11cases} there are also models with $D$-flat scalar field 
directions, a property which can be phenomenologically advantageous. There are
also solutions simultaneously in accordance with $E_8$ charge quantization (see
Eq.~(\ref{quant})) and the proton lifetime. 

I have further assumed that all exotic fields receive their masses through
couplings to MSSM singlets acquiring VEVs upon \uonep breaking. But exotic
leptons participating in $SU(2) \times U(1)_Y$ symmetry breaking will in 
general receive mass contributions of ${\cal O} (M_Z)$ which could conceivably 
by themselves be large enough to be in agreement with current limits. This 
seems difficult but is a possible loophole which I have ignored in this work. 

An implicit assumption has been made concerning the precise form of gauge
coupling unification. It is the often made assertion that gauge coupling 
unification works satisfactorily in the MSSM and should not be altered in 
leading (one-loop) order. Extra matter multiplets are then usually restricted 
to appear exclusively in complete $SU(5)$ representations. In the present work,
I have already relaxed this restriction and included {\em quasi-complete\/} 
$SU(5)$ representations. However, gauge coupling unification works only 
approximately, with the prediction for the strong coupling constant, 
$\alpha_s (M_Z)$, significantly higher than most observations and 
the unification scale at least one order of magnitude lower than the string 
scale. It is fair to ask whether some arbitrary (incomplete from 
the perspective of $SU(5)$) representation could improve the quality of 
unification. Indeed, by including a set of fields which contributes stronger to
the $SU(3)$ $\beta$-function than to the one of $SU(2)$ improves the prediction
for $\alpha_s (M_Z)$. Furthermore, by having the smallest contribution to 
the $U(1)_Y$ $\beta$-function, one can increase the unification scale. 
The minimal choice is the set $G + W + E^\pm$, contributing 
$\Delta\beta_3 = 3$, $\Delta\beta_2 = 2$, and $\Delta\beta_1 = 6/5$, 
respectively. Alternatively, one could add another ${\bf 5} + {\bf \ovl{5}}$ of
$SU(5)$ to this set, or choose a family plus a mirror-family of quarks, 
$Q + \ovl{Q}$, $U + \ovl{U}$, and $D + \ovl{D}$. Both cases correspond to 
$\Delta\beta_3 = 4$, $\Delta\beta_2 = 3$, and $\Delta\beta_1 = 11/5$. The case 
$\Delta\beta_3 = 5$ (and the same $\beta$-function differences) would yield 
unification in the non-perturbative domain, but is also conceivable. In any 
case, these choices yield $SU(2) \times U(1)_Y$ unification slightly above 
the reduced Planck scale of $2.4\times 10^{18}$~GeV. The prediction for 
$\alpha_s (M_Z)$ is now below typical observed values but the discrepancy is in
general neither much better nor much worse than in the MSSM. If it is found 
that higher order or threshold effects tend to {\em raise\/} the prediction for
$\alpha_s$, these scenarios may be of potential interest, but I have not 
considered them in the analysis of this paper. In this context it is amusing to
note that one would obtain a very good prediction for $\alpha_s$, as well as 
unification very close to the string scale, if these extra fields have 
intermediate scale masses near the geometric mean of the string and electroweak
scales. 

Even given these possible directions it is possible to draw some general
conclusions from the kind of analysis introduced in this work. Having left 
the framework of $E_6$ inspired \uonep models, anomaly cancellation tends to
provide $U(1)$ groups which have charges to the ordinary fermions just as 
the $U(1)$ subgroups of $E_6$. The charges of some of the extra fields 
typically differ, sometimes forbidding fast proton decay mediating couplings. 
A more thorough phenomenological investigation will be presented elsewhere.

\section*{Acknowledgement:}
It is a pleasure to thank Paul Langacker and Michael Pl\"umacher for many 
fruitful discussions. This work was supported in part by U.S. Department of 
Energy Grant EY--76--02--3071.

\end{document}